\documentclass[pra,twocolumn,tbtags,superscriptaddress,floatfix]{revtex4}
\usepackage{amssymb}
\usepackage{graphicx,amsmath}
\usepackage{bm}
\usepackage{times}
\begin{document}

\title{Influence of qubits' nonradiative decay into a common bath on the
transport properties of microwave photons}

\begin{abstract}
We consider the influence of nonradiative damping of qubits on the
microwave transport of photons, propagating in an open
one-dimensional microstrip line. Within the framework of the
formalism of a non-Hermitian Hamiltonian we obtained the
expressions for the transmission and reflection coefficients for
two qubits which explicitly account for the indirect interaction
between qubits due to nonradiative decay into common bath. It is
shown that this interaction leads to the results that are
significantly different from those already known.

\end{abstract}


\date{\today }
\author{Ya. S. Greenberg}\email{yakovgreenberg@yahoo.com}
\affiliation{Novosibirsk State Technical University, Novosibirsk,
Russia}
\author{A. N. Sultanov}
\affiliation{Novosibirsk State Technical University, Novosibirsk,
Russia}

%\classification{85.25.Dq,~ 85.25.Cp,~ 85.25.Hv,~ 84.40.Az}

 \maketitle

%%%%%%%%%%%%%%%%%%%%%%%%%%%%%%%%%%%%%%%%%%%%
%% MAINMATTER
%%%%%%%%%%%%%%%%%%%%%%%%%%%%%%%%%%%%%%%%%%%%
\section{Introduction}
Coherent interaction of solid-state qubits with microwave modes of
the coplanar waveguide is a field of extensive  theoretical and
experimental research (see the reviews \cite {Bul11, Roy17} and
references therein). Unlike the real atoms in an optical cavity,
the artificial atoms have significant nonradiative damping
channels when the energy of the excited state of a qubit radiates
not into a waveguide, but transmits to other degrees of freedom,
not related to the radiation. The calculation of microwave
transport with due account of the nonradiative damping of qubits
is of great importance from the point of manipulation and control
of qubits by means of a microwave field.

Traditionally, the qubits damping is assumed to be uncorrelated:
every qubit decays to its individual bath. From the formal point
of view, it corresponds to the addition to the qubit excitation
frequency $ \Omega $ of the imaginary quantity $ -i \gamma $,
where $ \gamma $ is the decay rate of the excited state of a qubit
into a nonradiative channel \cite {Rom09, Fang14, Zheng13,
Cheng17}. It seems to be reasonable for one qubit in a waveguide.
However, for two qubits their decay through a common nonradiative
channel leads to their indirect interaction, so that the damping
rates of individual qubits are not independent any more
\cite{Ojan07}. In this case, the expressions for reflection and
transmission coefficients are substantially modified, so that a
simple substitution $ \Omega \rightarrow \Omega-i \gamma $ is no
longer correct. In particular, as was indicated in
\cite{Gonzalez13} such simple replacement is not sufficient for
correct description of the entanglement between qubits.

%A simplest way is a direct addition to  In the second method the
%nonradiative damping is taken into account in the Markov
%approximation by using of the Lindblad operator in the equation
%for the density matrix \cite {Om10, Shen09}. Both these methods
%lead to the same result: a formal replacement of $ \Omega $ by $
%\Omega-i \gamma $ in the expressions for transmission and
%reflection coefficients.

More generally, the influence of a common bath on the decoherence
and relaxation in qubit systems has been studied in
\cite{Storcz03, Storcz05} for superconducting qubits and in
\cite{Vor03, Brand99} for quantum dots. Specially prepared common
bath can also be used for the creation of non-decaying entangled
states in multi-qubit systems \cite{Reiter13, Rod17}.

In the present work we study a microwave photon transport through
an open waveguide with imbedded two qubits at a distance $ d $
from each other. Qubits are characterized by their  rates of
spontaneous emission into a waveguide $ \Gamma_1, \Gamma_2 $, the
rates of nonradiative damping into local channels $\kappa_1,
\kappa_2$ and in a common bath $ \gamma_1, \gamma_2 $. The damping
in a common bath gives rise to off- diagonal elements in
Hamiltonian matrix which significantly influences the transport
characteristics of microwaves photons.

The calculation of transport coefficients is carried out in the
formalism of the effective non-Hermitian Hamiltonian, which has a
rather wide range of applicability in the study of various kinds
of open quantum systems (see the review articles \cite
{Auerbach11, Volya05} and numerous examples and references
therein). This formalism in application to photon transport
through one-dimensional chain of two-level systems is described in
detail in \cite {Greenberg15}, hence here we will only limit it to
a concise  summary.\\
\section{Formulation of the problem}
The Hamiltonian of two qubits interacting with a photon field in
an open one dimensional waveguide, is as follows \cite {com2}:

\begin{equation}\label{H}
\begin{array}{l}
H = \sum\limits_{i = 1,2} {\frac{1}{2}{\Omega _i}\left( {1 +
\sigma _Z^{(i)}} \right)
 + \sum\limits_k {{\omega _k}a_k^ + {a_k}} } \\ +
 J\left( {\sigma _ + ^{(1)}\sigma _ - ^{(2)} + \sigma _ + ^{(2)}\sigma _ - ^{(1)}} \right)\\
+ \sum\limits_k {\sum\limits_{i = 1,2} {{\lambda _i}\left( {a_k^ +
{e^{ - ik{x_i}}} + {a_k}{e^{ik{x_i}}}} \right)\sigma _x^{(i)}}
}+H_{\gamma}
\end{array}
\end{equation}

where $ \Omega_i $, $ (i = 1,2) $ is the qubit excitation
frequency, $ \omega_k $  is the waveguide modes, $ J $ is the
strength of the direct exchange interaction between qubits, $
\lambda_i $  is the interaction strength of the $ i $ -th qubit
located at the point $ x_i $ with a photon field, $ a_k ^ +, a_k $
are creation and annihilation photon operators, $ \sigma_ \alpha ^
{(i)} $ are the Pauli spin matrices, where a superscript refers to
the qubit number. Further, we take a coordinate origin in the
middle point between two qubits, so that $ x_1 = -d / 2 $, $ x_2 =
d / 2 $.

The fourth term in (\ref {H}) is responsible for spontaneous
emission of qubits into a waveguide, and the quantity  $ H _
{\gamma} $, which we do not specify here, is responsible for
nonradiative decay of the excited state of a qubit.

\section{Calculation of the effective non-Hermitian Hamiltonian}
According to the method of the effective non-Hermitian
Hamiltonian, the Hilbert space is divided into two mutually
orthogonal subspaces: the internal subspace $ Q $ and the external
subspace $ P $. The subspace $ Q $ describes a closed system of
the stationary states $ | n \rangle $ that does not interact with
the external environment. The subspace $ P $ contains, in addition
to the states of a closed system, also the states from the
continuum, to which  the system $ Q $ decays due to its
interaction with the system $ P $. Due to this interaction the
system $ Q $ becomes unstable: discrete energies acquire negative
imaginary components, which means in a time domain a decay of the
system $ Q $.

In our case the subspace $ Q $ consists of two vectors,
corresponding to the states in which one of the two qubits is in
the excited state $ | e \rangle $, and the other is in the ground
state $ | g \rangle $: $ | 1 \rangle = | e_1g_2 \rangle $, $ | 2
\rangle = | g_1e_2 \rangle $. The subspace $ P $ contains vectors
with both qubits being in the ground state and one photon being
either in the waveguide or in the nonradiative decay channel. The
dynamics of the entire system described by the Hamiltonian (\ref
{H}) can be project on the evolution of the system $ Q $ through
an effective the non-Hermitian Hamiltonian whose matrix elements
in the basis of the states of the system $ Q $ can be written as
follows \cite {Greenberg15}:

\begin{equation}\label{H1}
    \left\langle {m} \right|{H_{eff}}\left| {n} \right\rangle  =
\left\langle {m} \right|H\left| {n} \right\rangle
-\frac{i}{2}\sum\limits_{c = 1}^3 {A_m^c A_n^c } \nonumber
\end{equation}
\begin{equation}\label{H2}
+ \frac{1}{{2\pi }}\int\limits_{ - \infty }^{ + \infty }
{dq\frac{{{A_m}(q)A_n^*(q)}}{{k  - {q} + i\varepsilon }}}
\end{equation}

where the states $ | m \rangle, | n \rangle $ belong to the set $
| 1 \rangle, | 2 \rangle $, $ k $ is the wave vector of the photon
scattered in the waveguide. The summation in (\ref{H2}) runs over
three decay channels: two local channels ($c=1,2$) and a common
channel ($c=3$). The imaginary quantity $ i \epsilon $ allows to
avoid singularities in the integral and ensures divergent
scattering waves. The amplitudes $ A_m (q) $ are matrix elements
of the transition between states of the subspace $ Q $ and the
states of subspaces $ P $, where there is one photon in the
waveguide. The amplitudes $ A_m ^ {\gamma} $ describe the damping
of the excited state of the $ m $ -th qubit in a nonradiative
decay channels. For local channels the amplitudes
$A_1^{1},A_2^{2}$ are different from zero, while
$A_1^{2}=A_2^{1}=0$. For third common channel the amplitudes
$A_1^{3},A_2^{3}$ are also different from zero.

A direct calculation of the amplitudes $ A_m (q) $ and the
integral in (\ref {H2}) leads to the following result \cite
{Greenberg15}:

\begin{equation}\label{A}
    A_m(q)=\sqrt{\Gamma_m}e^{iqx_m}
\end{equation}

where $\Gamma_m$ is the rate of spontaneous emission of  $m$-th
qubit into a waveguide:
\begin{equation}\label{Gamma}
    \Gamma_m=\frac{L\lambda_m^2}{\hbar^2v_g},
\end{equation}
$L$-the waveguide length, $v_g$- the group velocity of photons in
a waveguide.
\begin{equation}\label{int}
\frac{1}{{2\pi }}\int\limits_{ - \infty }^{ + \infty }
{dq\frac{{{A_m}(q)A_n^*(q)}}{{k  - {q} + i\varepsilon
}}}=-i\sqrt{\Gamma_n\Gamma_m}e^{ik|d_{mn}|}
\end{equation}
where $d_{mn}=x_m-x_n$, $k$ is the wavevector which is related to
the frequency of scattered photon, $k=\omega/v_g$.

We consider the amplitudes $A_m^{\gamma}$ as constant quantities
which do not depend on the energy of the scattered photon:
$A_1^{1}=\sqrt{\kappa_1}$, $A_2^{2}=\sqrt{\kappa_2}$,
$A_1^{3}=\sqrt{\gamma_1}$, $A_2^{3}=\sqrt{\gamma_2}$. In addition,
we assume the amplitudes $A_1^3, A_2^3$ are independent on the
inter- qubit distance $d$ which corresponds to Markov
approximation: the interaction between qubits in a common bath is
non- retarded. In other words, the interaction wavelength in a
common channel is much larger than the inter- qubit distance.

 Hence, in the basis set $|1\rangle,|2\rangle$ we
obtain from (\ref{H1}) the matrix of effective Hamiltonian:

\begin{equation}\label{Heff}
{H_{eff}} = \left( {\begin{array}{*{20}{c}}
{{\Omega _1} - i{{\widetilde \Gamma }_1}}&\Lambda \\
\Lambda &{{\Omega _2} - i{{\widetilde \Gamma }_2}}
\end{array}} \right)
\end{equation}
where
\begin{equation}\label{Lambda}
    \Lambda=J- i\sqrt{{\Gamma _1}{\Gamma_2}}{e^{ikd}}
    -\frac{i}{2}\sqrt{{\gamma_1\gamma_2}}\quad,
\end{equation}

$\widetilde{\Gamma}_i=\Gamma_i+\frac{1}{2}\kappa_i+\frac{1}{2}\gamma_i$,
$i=1,2$.

The positions of resonances in a complex plane are determined by
the equation det$(\widetilde{\omega}-H_{eff})=0$ which gives the
following result:

\begin{eqnarray}\label{roots}
%\begin{flushleft}
\widetilde{ \omega}_{\pm}  = \frac{1}{2}(\Omega _1 + \Omega _2) -
i\frac{1}{2}(\widetilde{\Gamma }_1 + \widetilde{\Gamma}_2)\nonumber\\[3pt]
 \pm\frac{1}{2}\sqrt {{{\left( {{\Omega _1} - {\Omega _2}
 + i[{\widetilde{\Gamma }_2} - {\widetilde{\Gamma} _1}]} \right)}^2}+4\Lambda^2}
%\end{flushleft}
\end{eqnarray}
Off-diagonal elements in the matrix (\ref {Heff}), in addition to
the direct interactions $ J $ describe an indirect interaction $
\sqrt {\Gamma_1 \Gamma_2} e^{ikd} $ due to spontaneous radiation
of the excited qubits, and the interaction $ \sqrt {\gamma_1
\gamma_2} / 2 $, caused by nonradiative decay into a common
channel. We note that the contribution of spontaneous emission
depends on the frequency of the scattered photon $ \omega $ ($ k =
\omega / v_g $). As a result the position of the resonances (\ref
{roots}) in the complex plane depends on the frequency of the
scattered photon. This is a manifestation of the retardation
effect. Thus, in this formalism, the non-Markovian effects in
photon- qubit interaction are taken into account automatically.
Therefore, the position of the resonances on the real frequency
axis is determined, in general, by a nonlinear equation $ \omega =
\text {Re} [\widetilde {\omega} _ {\pm} (\omega)] $.

For identical qubits we obtain from (\ref{roots}):
\begin{equation}\label{Re}
    \texttt{Re}\omega_{\pm}=\Omega\pm J\pm\Gamma \sin{kd}
\end{equation}

\begin{equation}\label{Im}
\texttt{Im}\omega_{\pm}=-\Gamma (1\pm\cos{kd})-\frac{1}{2}\kappa
 -\frac{1\pm 1}{2}\gamma
\end{equation}

\begin{figure}
  \includegraphics[width=8 cm]{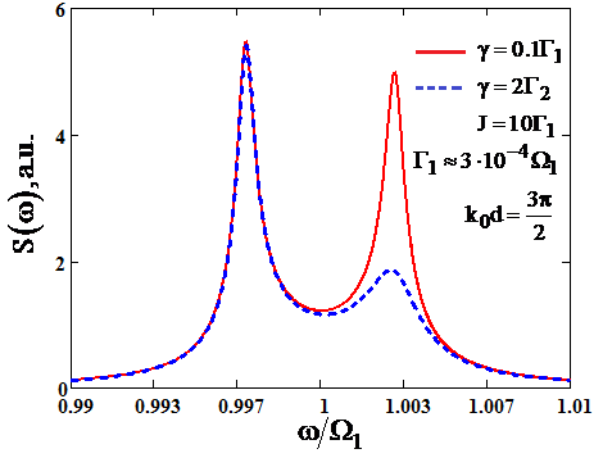}\\
  \caption{Resonance spectrum for two identical qubits
  depending on the value of nonradiative parameter $\gamma$.
   $k_0=\Omega/v_g$.}\label{w}
\end{figure}

As it follows from (\ref {Im}), the resonance width Im $ \omega _-
$ does not depend on the parameter $ \gamma $ and in the limit $
kd \ll 1 $ is defined only by relaxation $\kappa$ in a local
channel. This result is due to the interaction of non-radiative
decay into a common channel (the last term on the right hand side
of Eq. (\ref {Lambda})). If it had not been taken into account,
then the widths of both resonances would be proportional to $
\gamma $. Thus, with the increase of $ \gamma $ the width of one
resonance increases linearly, while the width of other one remains
unchanged. Fig. \ref {w} shows the resonance spectrum $ S (\omega)
$ for two identical qubits, which is determined by the zeros of
the real part of the determinant $ D (\omega) $ ($ S (\omega)
\approx 1 /D (\omega) $). The left peak corresponds to the
frequency Re $ \omega _- $, the right one to the frequency Re $
\omega _ + $. With the increase of $ \gamma $, the width and the
amplitude of the left peak remains unchanged while the width of
the right peak increases and its amplitude decreases. Here and
below the quantity $ k_0 = \Omega / v_g $.

\section{Two qubit entanglement}
The wave function for two qubit system interacting with a photon
field reads  (see Eq.42 in \cite{Greenberg15}):

\begin{equation}\label{Qb}
    \Psi_Q=\sum\limits_{n,m=1}^2 {\left| n \right\rangle }
\lambda_m R_{n,m}e^{ikx_m}
\end{equation}

In equation (\ref{Qb}) the matrix $R_{mn}$ is the inverse of the
matrix $(\omega-H_{eff})_{mn}$:
\begin{equation}\label{Rmn}
R = \frac{1}{{D(\omega )}}\left( {\begin{array}{*{20}{c}}
{\omega  - {\Omega _2} + i{{\widetilde \Gamma }_2}}&\Lambda \\
\Lambda &{\omega  - {\Omega _1} + i{{\widetilde \Gamma }_1}}
\end{array}} \right)
\end{equation}
where
\begin{equation}\label{det}
\begin{array}{l}
D(\omega ) = det(\omega  - {H_{eff}}) \\
 = (\omega  - {\Omega _1} + i{\widetilde \Gamma _1})(\omega  - {\Omega _2} + i{\widetilde \Gamma _2}) - {\Lambda ^2}\\
\end{array}
\end{equation}
Determinant (\ref{det}) can also be written as follows

\begin{equation}\label{det1}
    D(\omega)=(\omega-\widetilde{\omega}_-)(\omega-\widetilde{\omega}_+)
\end{equation}
where the complex roots $\widetilde{\omega}_{\pm}$ are determined
from (\ref{roots}).

From (\ref{Rmn}) we obtain the two qubit wavefunction:

\begin{equation}\label{psi_q}
 \begin{array}{l}
{\Psi _Q} = \frac{\lambda }{{D(\omega )}}(\omega  - \Omega  +
i\widetilde \Gamma ) \left( {\left| 1 \right\rangle {e^{ - ikd}} +
\left| 2 \right\rangle {e^{ikd}}} \right)\\[3pt]
 + \frac{\lambda }{{D(\omega )}}\Lambda
 \left( {\left| 1 \right\rangle {e^{ikd}}
 + \left| 2 \right\rangle {e^{ - ikd}}} \right)
\end{array}
\end{equation}

This expression allows us to determine the probability amplitudes
for the excitation of each of the qubits ($ \langle 1 | \Psi_Q
\rangle $, $ \langle 2 | \Psi_Q \rangle $), and the degree of
entanglement of two-qubit state. It follows from (\ref {psi_q})
that the maximally entanglement states $ \Psi_Q \approx (| 1
\rangle \pm | 2 \rangle) $ at arbitrary frequency $ \omega $ of
the scattered photon take place in the limit $ kd \ll 1 $, as well
as for discrete frequencies determined from the relation $ kd = n
\pi / 2 $, where $ n = 1, 2, 3 ..... $.

\begin{figure}
  \includegraphics[width=8 cm]{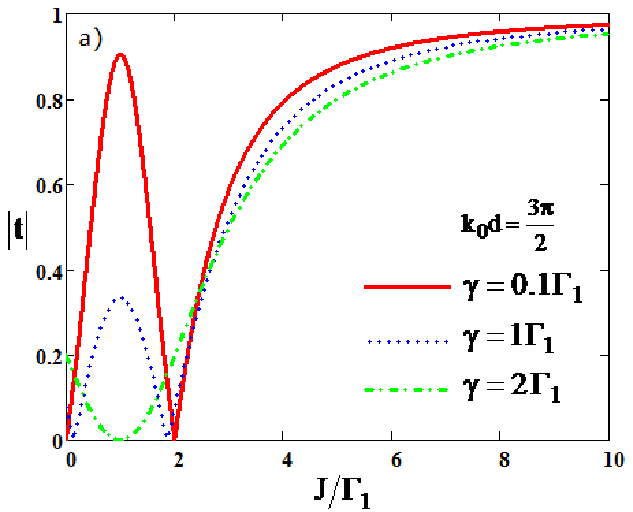}
  \includegraphics[width=8 cm]{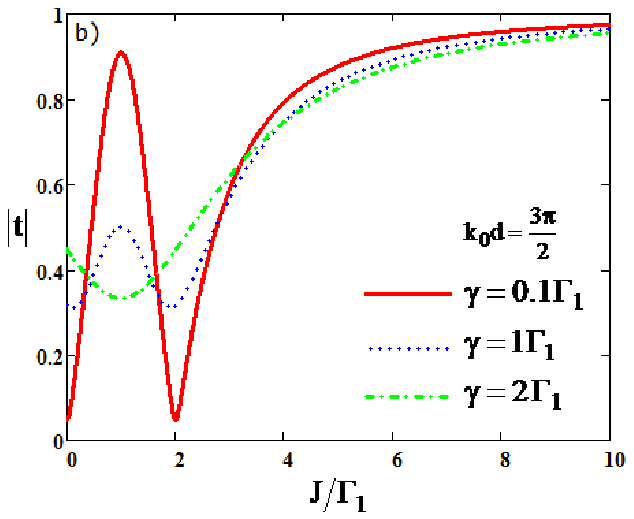}\\
  \caption{The dependence of transmission $|t|$
  on the direct coupling $J$ for different values of nonradiative decay parameter
  $\gamma$;
   a) calculation from equation (11a) in \cite{Cheng17}, b)
   calculation from equation
   (\ref{t1}).}\label{tr}
\end{figure}

\begin{figure}
  \includegraphics[width=8 cm]{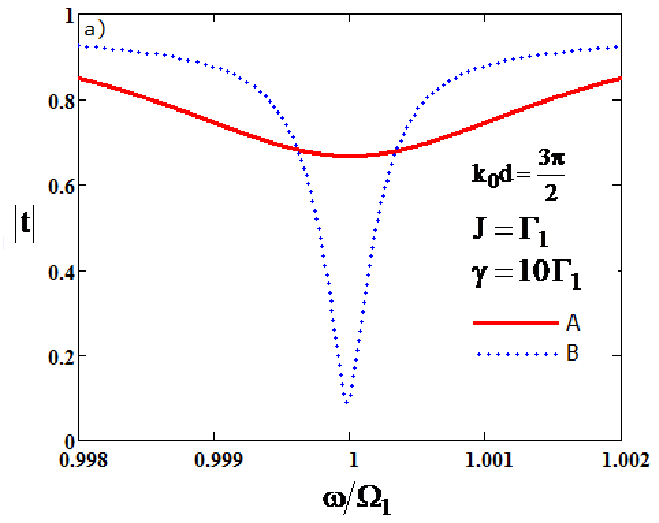}
\includegraphics[width=8 cm]{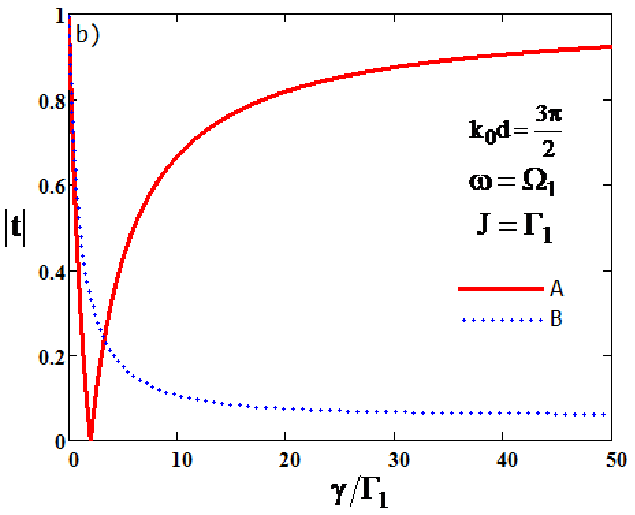}\\
  \caption{a) The dependence of transmission $|t|$
  on the frequency of scattered photon and b) on the nonradiative decay parameter
  $\gamma$; solid lines (A) correspond to equation (11a) from
  \cite{Cheng17}, dotted lines (B) are calculated from
   (\ref{t1}).}\label{reson}
\end{figure}

\begin{figure}
  \includegraphics[width=8 cm]{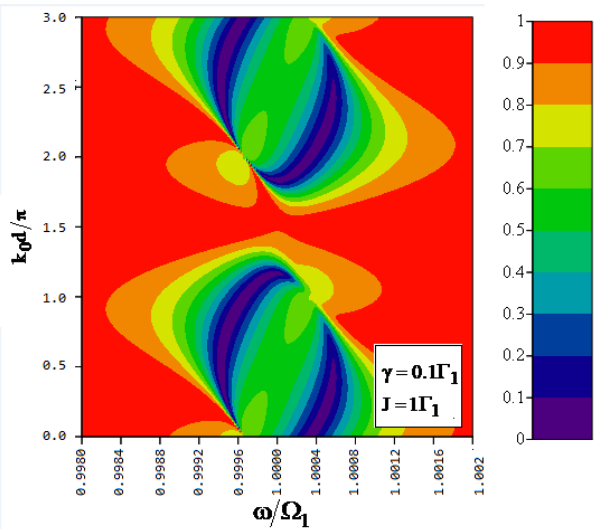}\\
  \caption{Two dimension (in the plain $kd, \omega$) pattern of transmission factor $|t|$
  Near $k_0d=3\pi/2$ the transparency region is well visible.}\label{trsp}
\end{figure}

\section{Calculation of the photon transport coefficients}
In accordance with the method of non Hermitian Hamiltonian the
photon transition from a state with an initial momentum $ k $ into
a state with a final momentum $ q $ is given by the amplitude
\cite{Volya05}:
\begin{equation}\label{Tab}
 {T^{qk}} = \sum\limits_{m,n = 1}^2 {A_m(q){R_{mn}}A_n^*(k)}
\end{equation}

Here the coefficients of photon transport $t^{qk}$ are the
elements of scattering matrix $S^{qk}$ which is related to
$T^{qk}$ as follows:

\begin{equation}\label{trans}
    t^{qk}\equiv S^{qk}=\delta_{qk}-iT^{qk}
\end{equation}

For further consideration we rewrite the determinant (\ref{det})
in the following form:

\begin{equation}\label{det1}
\begin{array}{l}
D(\omega ) = {\delta _1}{\delta _2} - {J^2} + i{\delta _1}{\Gamma _2} + i{\delta _2}{\Gamma _1} + 2iJ\sqrt {{\Gamma _1}{\Gamma _2}} {e^{ikd}}\\
 - {\Gamma _1}{\Gamma _2}\left( {1 - {e^{2ikd}}} \right) + A
\end{array}
\end{equation}
where
$\delta_j=\omega-\Omega_j+\frac{i}{2}\kappa_j+\frac{i}{2}\gamma_j,
j=1,2$.

In the last line of (\ref{det1}) we define the quantity $A$, which
is responsible for the interaction between qubits due to
nonradiative decay channels:
\begin{equation}\label{A1}
A = \frac{{\sqrt {{\gamma _1}{\gamma _2}} }}{4}\left( {\sqrt
{{\gamma _1}{\gamma _2}}  + 4i(J - i\sqrt {{\Gamma _1}{\Gamma _2}}
{e^{ikd}})} \right)
\end{equation}

The photon transmission corresponds to $q=k$, the reflection
corresponds to $q=-k$. Then, in accordance with (\ref{trans}) the
transmission $t$ and reflection $r$ coefficients can be expressed
as: $t\equiv t^{kk}=1-iT^{kk}$, $r\equiv t^{-kk}=-iT^{-kk}$. By
using (\ref{A}), (\ref{Tab}) and the explicit expression for $R$
matrix (\ref{Rmn}) we write these coefficients in the following
form:

\begin{equation}\label{t1}
t = \frac{1}{{D(\omega )}}\left( {{\delta _1}{\delta _2} - {J^2} -
2J\sqrt {{\Gamma _1}{\Gamma _2}} \sin kd + B} \right)
\end{equation}

\begin{equation}\label{r1}
\begin{array}{l}
r = \frac{{ - i}}{{D(\omega )}}\left[ {{\delta _2}{\Gamma
_1}{e^{ikd}}
 + {\delta _1}{\Gamma _2}{e^{ - ikd}} + 2{\Gamma _1}{\Gamma _2}\sin kd}
 \right.\\[3pt]
\quad\quad\left. { + 2J\sqrt {{\Gamma _1}{\Gamma _2}}+C } \right]
\end{array}
\end{equation}
where $D(\omega)$ was defined in (\ref{det1}), and the quantities
$B$ and $C$ as well as $A$ (\ref{A1}) describes the contribution
of nonradiative decay channels into the interaction between the
qubits:

\begin{equation}\label{B}
B = \frac{{\sqrt {{\gamma _1}{\gamma _2}} }}{4}\left( {\sqrt
{{\gamma _1}{\gamma _2}}  + 4i(J + \sqrt {{\Gamma _1}{\Gamma _2}}
\sin kd)} \right)
\end{equation}
\begin{equation}\label{C}
    C=-\sqrt{\Gamma_1\Gamma_2}\sqrt{\gamma_1\gamma_2}
\end{equation}

Neglecting the quantities $ A $, $ B $ and $ C $, in the
expressions (\ref {det1}), (\ref {t1}) and (\ref {r1}) we obtain
for the coefficients $ t $ and $ r $ the expressions which exactly
coincide with those already known (expressions (11a) and (11b) in
the work \cite {Cheng17}).

In Fig. \ref {tr} and Fig. \ref {reson} the transmission $ | t | $
calculated from (\ref {t1})
 is compared with the one calculated from the expression (11a) in\cite {Cheng17}).
The calculations were carried out for identical qubits: $ \Omega_1
= \Omega_2, \Gamma_1 = \Gamma_2, \gamma_1 = \gamma_2 $ for fixed
values $ \Omega_1 = 3 $ GHz, $ \Gamma_1 = 3 \times 10 ^ {- 4}
\Omega_1 $. From Fig. \ref {tr} it is clearly seen that in the
region $ J <2 \Gamma_1 $ the expression (\ref {t1}) leads to the
results that substantially differ from those which follow from the
expression (11a) in \cite {Cheng17}. Even more clearly this
difference is evident in Fig. \ref {reson}, where the presence of
indirect interaction between qubits due to nonradiative decay
channels results in a significant suppression of the photon
transmission. Two dimensional (in the plane $ k_0d, \omega $)
distribution of the transmission $|t|$ calculated from (\ref {t1})
is shown in Fig. \ref {trsp}. Near $ k_0d = 3 \pi / 2 $, the
region, where the transmission coefficient is close to unity in a
wide frequency range, is clearly visible. The numerical
simulations showed that the width of this region along the
vertical axis is proportional to the magnitude $ J $ and inversely
proportional to the rate of nonradiative decay $ \gamma $.

\section{Conclusion.}
It was shown that the existence of nonradiative decay into a
common bath results in additional interaction between qubits,
which renders significant influence on the transmission and
reflection of microwave photons. We studied in detail a two qubit
system in an open photonic waveguide. The analytic expressions for
the transmission and reflection coefficients were obtained. It is
shown that there are parameter regions, where the inter- qubit
correlations due to relaxation into a common bath lead to the
results which are significantly different from the known ones.

The work was supported by the Russian Science Foundation under
project 16-19-10069.


\begin{thebibliography}{9}
\bibitem{com2} Throughout the paper we put $\hbar=1$, so that all energies
 in (\ref{H}) and subsequent expressions are expressed in
 frequency units.
\end{thebibliography}
\end{document}